\documentclass[pre,aps,twocolumn,showpacs,floatfix]{revtex4}

\usepackage[dvips]{graphicx}
\usepackage{amsmath}
\usepackage{amssymb}
\usepackage[nice]{nicefrac}
\usepackage{color}

\newcommand{\Ld}{L_{\alpha,\beta}}
\newcommand{\Tm}{T_{\mathrm{max}}}
\newcommand{\Nb}{N_{\mathrm{bins}}}
\newcommand{\xtr}{x_{\mathrm{tr}}}
\newcommand{\Veff}{V_{\mathrm{eff}}(x)}

\begin{document}
\title{Stationary states in Langevin dynamics under asymmetric L\'evy noises}

\author{B. Dybiec}
\email{bartek@th.if.uj.edu.pl}
\affiliation{M. Smoluchowski Institute of Physics, and Mark Kac Center for Complex Systems Research, Jagellonian University, ul. Reymonta 4, 30--059 Krak\'ow, Poland}

\author{E. Gudowska-Nowak}
\email{gudowska@th.if.uj.edu.pl}
\affiliation{M. Smoluchowski Institute of Physics, and Mark Kac Center for Complex Systems Research, Jagellonian University, ul. Reymonta 4, 30--059 Krak\'ow, Poland}

\author{I. M. Sokolov}
\email{igor.sokolov@physik.hu-berlin.de}
\affiliation{Institut f\"ur Physik, Humboldt-Universit\"at zu Berlin, Newtonstrasse 15, D--12489 Berlin, Germany}

\date{\today}
\begin{abstract}
Properties of systems driven by white non-Gaussian noises can be very different from these of systems driven by the white Gaussian noise. We investigate stationary probability densities for systems driven by $\alpha$-stable L\'evy type noises, which provide natural extension to the Gaussian noise having however a new property mainly a possibility of being asymmetric. Stationary probability densities are examined for a particle moving in parabolic, quartic and in generic double well potential models subjected to the action of $\alpha$-stable noises. Relevant solutions are constructed by methods of stochastic dynamics. In situations where analytical results are known they are compared with numerical results. Furthermore, the problem of estimation of the parameters of stationary densities is investigated.
\end{abstract}

\maketitle

\section{Introduction\label{sec:introduction}}

Behavior of many natural systems in contact with their surroundings can be described within a
stochastic picture based on Langevin equations. The basic equation of this type reads
\begin{equation}
\dot{x}(t)=f(x)+\zeta(t),
\label{eq:langevin0}
\end{equation}
where $f(x)$ is the deterministic ``force'' representing the
internal dynamics of the system and $\zeta(t)$ is the ``noise''
describing its interaction with its complex surrounding (heat
bath). In many cases this noise can be considered as white and
Gaussian, giving rise to the classical Langevin approach used in
the analysis of Brownian motion. The whiteness of the noise (lack
of temporal correlations) corresponds to the existence of
time-scale separation between the dynamics of a relevant variable
of interest $x(t)$ and the typical timescale of the noise. White
noise can be thus considered as a standard stochastic process that
describes in the simplest fashion the effects of ``fast''
surrounding. On the other hand, the Gaussian nature of the noise
is usually guaranteed by assuming the surrounding bath being
composed  of many practically independent subsystems and by the
fact that the interaction of $x$ with each of these subsystems is
bounded. The first assumption allows for considering the noise as
being a sum of many independent random contributions (in
thermodynamical limit -- infinitely many), which mathematically
corresponds to the statement that its probability distribution is
infinitely divisible and stable. The second assumption chooses the
Gaussian distribution as the only one possessing finite
dispersion. However, the assumption that the perturbations in the
system's dynamics due to interactions with bath are described by
white Gaussian noise is not always appropriate when describing
real processes where each of the assumptions concerning the noise (e.g.
its whiteness or its Gaussian distribution) can be violated. In various
phenomena in physics, chemistry or biology
\cite{schlesinger1995,nielsen2001} the noise can be still interpreted as
white (i.e., with stationary, independent increments) and
distributed according to a stable and infinitely divisible law,
however, the distribution of the noise variable $\zeta$ is
registered as following not a Gaussian, but rather a more general,
L\'evy probability distribution. Such situations were addressed
for example in Refs.
\cite{jespersen1999,ditlevsen1999,kosko2001,dybiec2004,dybiec2004b,dybiec2004c,dybiec2006,dybiec2006b,dybiec2007,chechkin2002,chechkin2003,chechkin2004,sokolov2003,dybiecphd}.
The present work discusses some further properties of L\'evy
flights in external potentials with a focus on astonishing aspects
of noise-induced bifurcations and explores in more detail features
of stationary states in Langevin systems under the influence of
asymmetric L\'evy noises.

L\'evy distributions $L_{\alpha,\beta}(\zeta;\sigma,\mu)$ correspond to a 4-parametrical family of
the probability density functions characterized by their Fourier-transforms (characteristic functions
of the distributions)
$\phi(k) = \int_{-\infty}^\infty e^{ik\zeta} L_{\alpha,\beta}(\zeta;\sigma,\mu) d\zeta$ being \cite{feller1968,janicki1994,janicki1996}
\begin{equation}
\phi(k) = \exp\left[ ik\mu -\sigma^\alpha|k|^\alpha\left( 1-i\beta\mbox{sgn}(k)
\tan\frac{\pi\alpha}{2} \right)\right]
\label{eq:charakt}
\end{equation}
for $\alpha\in(0,1)\cup(1,2]$ and
\begin{equation}
\phi(k) = \exp\left[ ik\mu -\sigma|k|\left( 1+i\beta\frac{2}{\pi}\mbox{sgn} (k) \ln|k| \right) \right]
\label{eq:charakt1}
\end{equation}
for $\alpha=1$ \cite{feller1968,janicki1994,janicki1996}. Here the
parameter $\alpha$ (where $\alpha\in(0,2]$) is the stability index
of the distribution describing (for $\alpha <2$) its asymptotic
``fat'' tail characteristics yielding  $L_{\alpha,\beta}(\zeta;
\sigma,\mu)\sim |\zeta|^{-(1+\alpha)}$ for large $\zeta$.
 The parameter $\sigma$ characterizes a scale and  $\beta\in[-1,1]$
 defines a skewness (asymmetry) of the distribution, whereas $\mu$
 denotes the location parameter.
 As it is clear from Eq.~(\ref{eq:charakt}), Gaussian
distribution corresponds to a special case of a L\'evy law for
$\alpha=2$, with $\mu$ interpreted now as a mean and $\sigma$ as
the dispersion of the distribution. However, this special case is
somewhat degenerated: the dependence of the distribution on
$\beta$ disappears due to the fact that $\tan \pi =0$, so that all
Gaussian distributions are symmetric! In general cases strongly
asymmetric distributions (up to extreme, one-sided ones) may
appear, see e.g. \cite{sokolov2003,eliazar2003,eliazar2004}. Such
realms are discussed much less extensively than the cases of
symmetric noises
\cite{chechkin2002,chechkin2003,chechkin2004,ditlevsen1999,jespersen1999}.

In many situations, one is interested not in the individual
properties of the trajectories $x(t)$ but, instead, in the
one-point probability distributions defined on ensemble of
trajectories: $P(x,t)=\langle \delta(x-x(t))\rangle_\zeta$ at some
given time $t$. For stochastic system described by the Langevin
equation with an additive  white L\'evy noise forcing, the
distribution function $P(x,t)$ of the variable $x(t)$ fulfills the
associated fractional differential Fokker-Planck equation.
Stationary states (if they exist) can be  then read from the
asymptotic time independent solutions $P(x)= \lim_{t \rightarrow
\infty} P(x,t)$. Such solutions were discussed e.g. in
\cite{jespersen1999,chechkin2002,chechkin2003,chechkin2004} where
analysis of L\'evy flights in harmonic/superharmonic potentials
has been presented. Nevertheless, the discussion presented there
is far from being complete, since only symmetric L\'evy
distributions have been considered. In the present work we pay
special attention to asymmetric ones.

In this article we focus on stationary states for a particle moving in
quadratic, quartic and double well potentials subjected to $\alpha$-stable white
noises. Theoretical descriptions of such systems is based on the Langevin
equation and/or Fokker-Planck equation, which in general is of the fractional
order \cite{podlubny1998}. The model under discussion is presented in
Section~\ref{sec:model}. Section~\ref{sec:stationary} discusses obtained
results, which are divided into three subsections regarding results for
parabolic and quartic potential (Sections~\ref{sec:parabolic} and
\ref{sec:quartic}) and double well potential model
(Section~\ref{sec:doublewell}). The paper is closed with concluding remarks
(Section~\ref{sec:summary}). Additional information, regarding problem of the dimensionality of the Langevin equation is included in Appendix~\ref{sec:appendix}.

\section{Model\label{sec:model}}

\begin{figure}
\begin{center}
\includegraphics[angle=0,width=8cm]{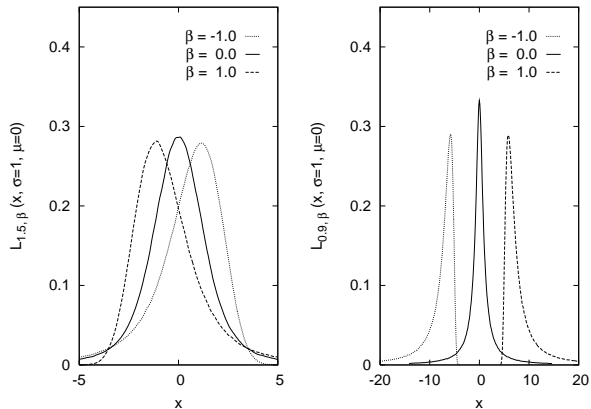}
\caption{Sample $\alpha$-stable probability density functions
(PDF) with $\alpha = 1.5$ (left panel) and $\alpha = 0.9$ (right
panel). For $\beta = 0$ distributions are symmetric, while for
$\beta=\pm 1$ they are asymmetric functions. The support of PDFs
for the fully asymmetric cases with $\beta=\pm 1$ and $\alpha < 1$
(right panel) assumes only negative values for $\beta = -1$ and
only positive values for $\beta = 1$. Note the differences in the
positions of the maxima for $\alpha<1$ and $\alpha>1$.}
\label{fig:density}
\end{center}
\end{figure}

Let us consider a motion of an overdamped particle in a field of a potential force, so that Eq.~(\ref{eq:langevin0}) takes the form of
\begin{equation}
\dot{x}(t)=-V'(x)+\zeta(t),
\label{eq:langevin}
\end{equation}
and $\zeta(t)$ denotes a L\'evy stable white noise process
\cite{weiss1983,caceras1999,jespersen1999,ditlevsen1999,dybiec2006}. The
value of the stochastic process defined by Eq.~(\ref{eq:langevin})
can be calculated as \cite{janicki1994,janicki1994b}
\begin{eqnarray}
x(t)& = & x(0)-\int\limits_0^t V(x(s))ds + \int\limits_0^t\zeta(s)ds \nonumber \\
& = & x(0)-\int\limits_0^t V(x(s))ds + \int\limits_0^t d\Ld.
\end{eqnarray}
Here, the integral $\int_0^t\zeta(s)ds\equiv \int_0^t d\Ld$
defines a generalized Wiener process
\cite{weiss1983,jespersen1999,ditlevsen1999,dybiec2006} that is driven
by a L\'evy stable noise, whose increments are distributed
according to a stable density with the index $\alpha$.
The L\'evy noise is a formal time derivative of the generalized Wiener process. For the time step of integration $\Delta t$, the increments of the generalized Wiener process are distributed according to the distribution
$L_{\alpha,\beta}(\Delta x;\sigma(\Delta t)^{1/\alpha},\mu=0)$
\cite{janicki1994,janicki1996,janicki2001,nolan2002}. We discuss
the overall range of parameters $\alpha \in (0,2]; \; \beta \in
[-1,1]$ excluding the case of $\alpha=1$ with $\beta\neq0$, for
which the numerical results are unreliable due to well known
numerical instabilities
\cite{janicki1994,weron1995,janicki1996,dybiec2004}. Putting the
location parameter of the distribution to zero does not influence
the generality of our results: Taking location parameter $\mu$ to
be nonzero is equivalent to adding a linear term to the potential
(constant drift). Sample $\alpha$-stable probability densities are
presented in Fig.~\ref{fig:density}.

The Langevin equation~(\ref{eq:langevin}) describes evolution of a single
realization of the stochastic process $\{ x(t) \}$. Random numbers distributed
according to a canonical form of characteristics functions given by
Eqs.~(\ref{eq:charakt})--(\ref{eq:charakt1}) can be generated using the Janicki-Weron algorithm \cite{weron1995,weron1996}. More details on numerical scheme of
integration of stochastic differential equations with respect to $\alpha$-stable
noises can be found elsewhere
\cite{janicki1994,janicki1996,janicki2001,dybiec2004b,dybiec2006}.

For $\alpha\neq 1$, equation (\ref{eq:langevin}) is associated
with the following fractional Fokker-Planck equation (FFPE)
\cite{metzler1999,yanovsky2000,schertzer2001,paola2003,brockmann2002}
\begin{eqnarray}
\label{eq:ffpe}
\frac{\partial P(x,t)}{\partial t} & = & -\frac{\partial}{\partial x}\left[ \mu - V'(x,t) \right]P(x,t) \\
& + & \sigma^\alpha\left[ \frac{\partial^\alpha}{\partial |x|^\alpha} P(x,t) \right. \nonumber \\
& + & \left. \beta\tan\frac{\pi\alpha}{2}\frac{\partial}{\partial
x} \frac{\partial^{\alpha-1}}{\partial |x|^{\alpha-1}} P(x,t)
\right], \nonumber \label{ffpe}
\end{eqnarray}
where the fractional (Riesz-Weyl) derivative can be understood in
the sense of the Fourier transform
\cite{jespersen1999,chechkin2002,metzler2004,chechkin2003}
$\frac{\partial^\alpha}{\partial|x|^\alpha}f(x)=-\int_{-\infty}^{\infty}\frac{dk}{2\pi}e^{-ikx}|k|^\alpha
\hat{f}(k).$ The fractional derivatives in (\ref{eq:ffpe})
originate from the form of the characteristic function, see
Eqs.~(\ref{eq:charakt}) and (\ref{eq:charakt1}), of L\'evy stable
variables \cite{schlesinger1995,yanovsky2000,paola2003,dubkov2005b}. The
nonzero asymmetry leads to an additional, asymmetric diffusion
term including an even, reflection-invariant Riesz-Weyl operator
and an odd first derivative which changes its sign under the $x
\rightarrow -x$ transformation. The overall order of derivatives
in the diffusion terms is the same, namely $\alpha$.

In the following, the value of the location parameter, $\mu$, is set to zero,
which guarantees that for $\alpha\neq 1$ L\'evy noise present in
Eq.~(\ref{eq:langevin}) is strictly stable and standard numerical methods of
integration of stochastic differential equations with respect to $\alpha$-stable
noises (namely the generalization of the Euler scheme) can be used
\cite{janicki1994,janicki1996,janicki2001}.

\begin{figure}
\begin{center}
\includegraphics[angle=0,width=8cm]{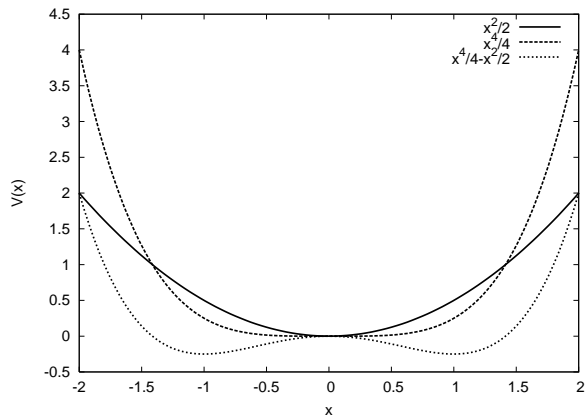}
\caption{Exemplary shapes of potential wells used in the study.
The examination of stationary states of the system perturbed by
$\alpha$-stable L\'evy type noises has been performed for a
generic double well potential model ($V(x)=x^4/4-x^2/2$), as well
as for  parabolic ($V(x)=x^2/2$) and quartic ($V(x)=x^4/4$)
potentials. As discussed in the text, the confinement of
trajectories (observation of bounded states) for $\alpha<2$  is
possible only if the potential slopes are steeper than for a
harmonic case. } \label{fig:setup}
\end{center}
\end{figure}

The behavior of a system where a particle is subject to the additive, white,
strongly non-Gaussian noise could be very different from the behavior in the
Gaussian regime \cite{dybiecphd}. In the Gaussian case, any potential well such
that $V(x)\to+\infty$ for $|x|\to+\infty$, even the piecewise linear one, is
sufficient to produce bounded states, i.e., the ones with a finite dispersion of
the particle's position. On the contrary, for the L\'evy noises with $\alpha<2$,
the potential which grows faster than quadratically in $x$ is required to
produce bounded states. Furthermore, in the Gaussian case stationary probability
distributions for a single-well potential are unimodal, which is no always true
for a L\'evy stable noise with the stability index $\alpha<2$
\cite{chechkin2002,chechkin2003}. Qualitative and quantitative differences are
caused by the fact that stable distributions are heavy-tailed and allow larger
noise pulses with a higher probability than the Gaussian distribution
\cite{nolan2002}. Moreover, stationary probability distributions for the
additive L\'evy noises ($\alpha<2$), if they exist, are not of
the Boltz\-mann-Gibbs type \cite{chechkin2002,chechkin2003,eliazar2003,barkai2003,barkai2004}.

In the following sections properties of stationary probability distributions for
systems perturbed by the general L\'evy noises are discussed. The performed
simulations corroborate earlier theoretical findings
\cite{jespersen1999,chechkin2002,chechkin2003,chechkin2004}. Furthermore, the
influence of the nonzero asymmetry parameter $\beta$ on the shape of stationary
distributions is discussed.

\section{Stationary States for a ``L\'evy-Brownian'' Particle\label{sec:stationary}}

The stationary probability densities $P(x)$ can be obtained either
by analytically solving Eq.~(\ref{eq:ffpe}) (which is
unfortunately possible only for a quite restricted set of special
cases) or, otherwise,  numerically. In such a case, there are two
approaches possible: either using the discretization of
Eq.~(\ref{eq:ffpe})
\cite{press1992,chechkin2002,chechkin2004,abdelrehim} or employing
a Monte-Carlo method based on simulation of
Eq.~(\ref{eq:langevin}),
\cite{janicki1994,janicki1994b,janicki1996,janicki2001}.
For the Gaussian noise the solutions of the Fokker-Planck
equation can be readily obtained by using shooting methods and
discretization techniques \cite{dybiec2007c}. For the general
L\'evy case such solutions can be constructed by discretization of
Eq.~(\ref{eq:ffpe}), which converts partial differential equation
to a discrete Markov chain
\cite{gorenflo2002,abdelrehim}. However, this approach
has a drawback of slow convergence and possible instability for
$\alpha<2$ \cite{meerschaert2004}, and consequently was not used
here. Thus, our data are based on Monte-Carlo simulations; our
method of solution of the Langevin equation is based on the
slightly modified standard integration scheme for stochastic
equations of type (\ref{eq:langevin}) driven by $\alpha$-stable
L\'evy type noises \cite{janicki1994,janicki1996,janicki2001}, see
below. Stationary PDFs were extracted from ensembles of,
typically, $N=10^6$ trajectories of a given length $\Tm=10$. The
value of $\Tm$ was chosen by trial and comparison of numerical
estimates of $P(x, \Tm)$ for various $\Tm$ (sufficiently long
times $\Tm$ are requited to let $P(x,\Tm)$ reach stationarity).
A problem related to the choice of the simulation time $\Tm$ is the choice of the time step of integration $\Delta t$. The simulations have been performed with the time step of integration $\Delta t=10^{-3}$. Such a choice of $\Delta t$ guarantees a compromise between accuracy and the computational cost of simulations. It is also suggested by earlier studies \cite{dybiecphd,dybiec2004,dybiec2004b,dybiec2006,dybiec2007}. Furthermore, $\Delta t=10^{-3}$ makes the $x$-domain in which the generalized Euler scheme can be used sufficiently large, see Section \ref{sec:quartic}.

\begin{figure}
\begin{center}
\includegraphics[angle=0,width=8cm]{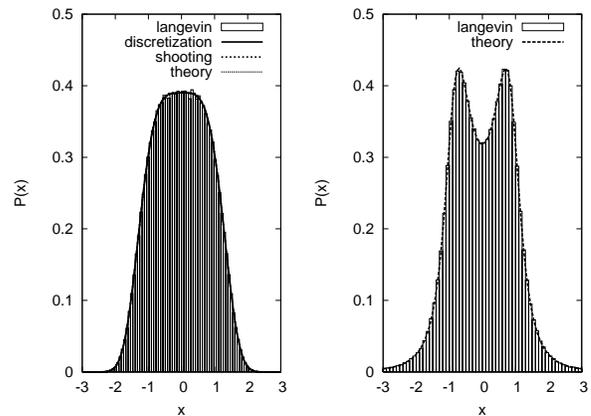}
\caption{Stationary solutions to the Smoluchowski-Fokker-Planck
and to the fractional Fokker-Planck equations associated with the
diffusion ($\alpha=2$, left panel) and  superdiffusion
($\alpha=1$, right panel) in a quartic potential $V(x)=x^4/4$. For
$\alpha=2$ numerical results were constructed by discretization
techniques and shooting methods \cite{dybiec2007c}. Simulation
parameters: $N=10^6$, $\Delta t=10^{-3}$, $\Nb=100$ and $\Tm=10$.}
\label{fig:static}
\end{center}
\end{figure}

For $\alpha=2$ (and an arbitrary skewness parameter $\beta$), the
random force term in the Langevin equation (\ref{eq:langevin})
represents a Gaussian white noise $\langle \zeta(t)\zeta(s)
\rangle _{\alpha=2}=2\delta(t-s)$ and the associated
Smoluchowski-Fokker-Planck equation governing evolution of the
probability density $P(x,t)$ reads
\begin{equation}
\frac{\partial P(x,t)}{\partial t} = \frac{\partial}{\partial x} V'(x)P(x,t) +\sigma^2 \frac{\partial^2}{\partial x^2} P(x,t),
\label{eq:fpe}
\end{equation}
with a stationary solution assuming the standard
Boltz\-mann-Gibbs form
\begin{equation}
P(x)=\mathcal{N}\exp\left[-\frac{V(x)}{\sigma^2} \right],
\label{eq:gibbspdf}
\end{equation}
of finite mean and variance. In contrast, L\'evy flights in
external potentials exhibit unexpected properties
\cite{jespersen1999,chechkin2002,chechkin2004,dubkov2005b} and their
stationary PDFs can be shown to possess  finite first and second
moments only, if the imposed deterministic forces are derived from
steeper than the parabolic potentials, see Fig.~\ref{fig:density}.
  As an example, in Fig.~\ref{fig:static}
stationary PDFs for a particle moving in the quartic potential
subject to the white Gaussian noise (left panel) and white Cauchy
noise (right panel) are compared. Numerical results were
constructed from the ensemble of final positions reached after the
long time $\Tm$ obtained from the simulation of the Langevin
equation (\ref{eq:langevin}) with $\alpha=2$, $\alpha=1$ and
$\sigma^2=1$, cf. Section~\ref{sec:quartic}. In forthcoming Sections we
are addressing this point by investigating Langevin dynamics
driven by asymmetric L\'evy white noises.

\subsection{Parabolic potential and algorithm testing\label{sec:parabolic}}

For $\sigma=1$, the fractional Fokker-Planck equation
(\ref{ffpe}) can be rewritten in the Fourier space in the form of
\begin{equation}
\frac{\partial \hat{P}(k,t)}{\partial t} = \hat{\mathbf{V}}
\hat{P}(k,t) - |k|^\alpha \hat{P}(k,t),
\end{equation}
where $\hat{\mathbf{V}}$ is the operator giving the Fourier
representation of the potential, which can be found in a closed
form only in the simplest cases, e.g. for polynomial potentials.
In the case of asymmetric distribution the analogous equation
reads
\begin{equation}
\frac{\partial \hat{P}(k,t)}{\partial t} = \hat{\mathbf{V}}
\hat{P}(k,t) - |k|^\alpha \left[1-i\beta\mbox{sign}(k)
\tan\frac{\pi\alpha}{2} \right] \hat{P}(k,t).
\end{equation}
The choice of the parabolic potential $V(x)=x^2/2$, see Fig.~\ref{fig:density}, results in its
Fourier transform  $\hat{\mathbf{V}}=-k\frac{\partial}{\partial
k}$. For symmetric $\alpha$-stable noises, the corresponding
equation for the stationary PDFs is
\cite{chechkin2002,chechkin2003}
\begin{equation}
\frac{\partial\hat{P}(k)}{\partial k}=-\mathrm{sign}k|k|^{\alpha-1}\hat{P}(k),
\end{equation}
and its solution in the Fourier space reads
\begin{equation}
\hat{P}(k)=\exp\left(-\frac{|k|^\alpha}{\alpha}\right),
\label{eq:parabolic}
\end{equation}
i.e., the stationary solution is a symmetric L\'evy distribution,
see Eqs.~(\ref{eq:charakt}) and (\ref{eq:charakt1}). Consequently,
the variance of the stationary solution is infinite. Therefore,
the parabolic potential is not sufficient to produce bounded
states for a particle subject to the action of a L\'evy noise
\cite{jespersen1999,chechkin2002,chechkin2003,chechkin2004,dubkov2005b}.
 For potentials steeper than the
parabolic well, the confinement of superdiffusive trajectories
becomes possible but, additionally, the additive L\'evy white
noise could induce bimodality in the stationary PDF.

For general $\alpha$-stable driving ($\alpha\neq 1$) the stationary solutions obey the equation
\begin{equation}
\frac{\partial \hat{P}(k)}{\partial k} = - \mathrm{sign}k |k|^{\alpha-1} \hat{P}(k) + i\beta \tan\frac{\pi\alpha}{2} |k|^{\alpha-1} \hat{P}(k),
\end{equation}
and its solutions is
\begin{equation}
\hat{P}(k)= \exp\left[ -\frac{|k|^\alpha}{\alpha}\left( 1-i\beta\mbox{sign}(k)
\tan\frac{\pi\alpha}{2} \right)\right].
\end{equation}
For the nonzero asymmetry parameter, $\beta$, like for the symmetric noise, the stationary probability density function is a stable law
with the same stability index $\alpha$ and the asymmetry parameter $\beta$ and a different scale parameter $\sigma'=\sigma \alpha^{-1/\alpha}$
(here $\sigma'=\alpha^{-1/\alpha}$). The location parameter $\mu$ of the resulting distribution is zero.
The existence of these analytical results allows to use the case of parabolic potentials as a test bench for our simulation algorithms.

Thus, for the testing purposes, large samples of long realizations of the
stochastic process given by Eq.~(\ref{eq:langevin}) were constructed. Using
these samples the values of the distributions parameters have been estimated
applying special software \cite{stable}. Estimated values of distributions
parameters are in good agreement with theoretical values, see
Tabs.~\ref{table4}--\ref{table6}. Techniques of estimation of the stable
distribution parameters are based on the evaluation of quantiles and
characteristic functions, and on maximum likelihood methods
\cite{marinelli2000} or direct use of time series \cite{siegert2001}. Results obtained by quantile methods and characteristic
function estimation seems to be more consistent with theoretical values than
results following from maximum likelihood, see Tabs.~\ref{table4}--\ref{table6}.
The largest differences between theoretical and estimated values of parameters
are observed for the location parameter $\mu$. In some situations, marked with
$*$, the program used \cite{stable} warns about numerical problems in evaluation
of the distributions parameters. To check whether results are influenced by the
length of simulation results for $\Tm=10$ and $\Tm=15$ were compared. Estimated
values of parameters for both values of $\Tm$ are consistent. Therefore, only
results for $\Tm=15$ are presented, see Tabs.~\ref{table4}--\ref{table6}.

\begin{table}[!h]
\begin{footnotesize}
\begin{center}
\begin{tabular}{c|c|c|c}
$\alpha$ & $\beta$ & $\sigma$ & $\mu$ \\ \hline

\textbf{0.5} & \textbf{-1} & \textbf{4} & \textbf{0}\\
0.502 & -0.964 & 4.186 & -5.43$\times 10^{-2}$\\
0.499 & -1.000 & 3.978 & -1.39$\times 10^{-2}$\\
*0.500 & -0.997 & 2.541 & -6.81$\times 10^{-2}$\\ \hline

\textbf{0.5} & \textbf{-0.5} & \textbf{4} & \textbf{0}\\
0.500 & -0.500 & 3.986 & -1.76$\times 10^{-3}$\\
0.500 & -0.498 & 3.992 & -1.08$\times 10^{-2}$\\
0.514 & -0.499 & 4.023 & 6.19$\times 10^{-2}$\\ \hline

\textbf{0.5} & \textbf{0} & \textbf{4} & \textbf{0}\\
0.500 & 0.001 & 3.993 & -6.16$\times 10^{-3}$\\
0.501 & 0.000 & 3.998 & 6.24$\times 10^{-3}$\\
0.515 & 0.000 & 4.055 & -1.68$\times 10^{-3}$\\ \hline

\textbf{0.5} & \textbf{0.5} & \textbf{4} & \textbf{0}\\
0.502 & 0.503 & 3.988 & -3.00$\times 10^{-2}$\\
0.500 & 0.502 & 3.979 & -1.35$\times 10^{-2}$\\
0.515 & 0.500 & 4.020 & -6.52$\times 10^{-2}$\\ \hline

\textbf{0.5} & \textbf{1} & \textbf{4} & \textbf{0}\\
0.503 & 0.981 & 4.104 & -2.04$\times 10^{-2}$\\
0.500 & 1.000 & 3.986 & 5.90$\times 10^{-4}$\\
*0.462 & 0.990 & 7.844 & -0.19\\
\end{tabular}
\end{center}
\caption{Theoretical and estimated values of stationary PDF parameters. In every cell numbers in bold indicate theoretical values of parameters, followings rows estimated parameters using quantile evaluation (2$^\mathrm{nd}$ row), characteristic function evaluation (3$^\mathrm{rd}$ row) and maximum likelihood method (4$^\mathrm{th}$ row). Distributions' parameters were estimated by use of \cite{stable}. Simulations' parameters: $\Tm=15$, $\Delta t=10^{-3}$. Samples contain not less than $10^6$ elements. $*$ indicates cases when the software applied warned about some problems in estimation of sample parameters.}
\label{table4}
\end{footnotesize}
\end{table}

\begin{table}[!h]
\begin{footnotesize}
\begin{center}
\begin{tabular}{c|c|c|c}
$\alpha$ & $\beta$ & $\sigma$ & $\mu$ \\ \hline

\textbf{1.1} & \textbf{-1} & \textbf{0.92} & \textbf{0}\\
1.096 & -1.000 & 1.237 & -2.82\\
1.097 & -1.000 & 0.917 & -0.2\\
*1.100 & -1.000 & 0.917 & 2.70$\times 10^{-4}$\\ \hline

\textbf{1.1} & \textbf{-0.5} & \textbf{0.92} & \textbf{0}\\
1.099 & -0.505 & 0.916 & -4.15$\times 10^{-2}$\\
1.101 & -0.501 & 0.917 & 1.58$\times 10^{-2}$\\
1.100 & -0.501 & 0.917 & -5.55$\times 10^{-3}$\\ \hline

\textbf{1.1} & \textbf{0} & \textbf{0.92} & \textbf{0}\\
1.098 & -0.001 & 0.917 & -7.83$\times 10^{-3}$\\
1.098 & 0.002 & 0.916 & 8.93$\times 10^{-3}$\\
1.099 & -0.001 & 0.917 & -5.05$\times 10^{-3}$\\ \hline

\textbf{1.1} & \textbf{0.5} & \textbf{0.92} & \textbf{0}\\
1.098 & 0.497 & 0.916 & 4.00$\times 10^{-2}$\\
1.099 & 0.496 & 0.917 & 1.84$\times 10^{-2}$\\
1.099 & 0.496 & 0.917 & 2.33$\times 10^{-2}$\\ \hline

\textbf{1.1} & \textbf{1} & \textbf{0.92} & \textbf{0}\\
1.101 & 1.000 & 1.237 & 2.41 \\
1.100 & 1.000 & 0.918 & -1.68$\times 10^{-2}$\\
*1.100 & 1.000 & 0.918 & 4.17$\times 10^{-3}$\\
\end{tabular}
\end{center}
\caption{Continuation of Table~\ref{table4} ($\alpha=1.1$). }
\label{table5}
\end{footnotesize}
\end{table}

\begin{table}[!h]
\begin{footnotesize}
\begin{center}
\begin{tabular}{c|c|c|c}
$\alpha$ & $\beta$ & $\sigma$ & $\mu$ \\ \hline

\textbf{1.8} & \textbf{-1} & \textbf{0.72} & \textbf{0}\\
*1.798 & -0.994 & 0.721 & 1.25$\times 10^{-3}$\\
1.800 & -1.000 & 0.722 & 1.54$\times 10^{-3}$\\
1.829 & -0.990 & 0.723 & 2.03$\times 10^{-2}$\\ \hline

\textbf{1.8} & \textbf{-0.5} & \textbf{0.72} & \textbf{0}\\
1.806 & -0.531 & 0.723 & -2.02$\times 10^{-3}$\\
1.799 & -0.498 & 0.722 & -4.49$\times 10^{-4}$\\
1.849 & -0.545 & 0.729 & 1.17$\times 10^{-2}$\\ \hline

\textbf{1.8} & \textbf{0} & \textbf{0.72} & \textbf{0}\\
1.800 & -0.002 & 0.722 & -1.08$\times 10^{-3}$\\
1.798 & -0.001 & 0.721 & -5.69$\times 10^{-4}$\\
1.838 & 0.001 & 0.727 & -2.69$\times 10^{-4}$\\ \hline

\textbf{1.8} & \textbf{0.5} & \textbf{0.72} & \textbf{0}\\
1.799 & 0.491 & 0.722 & -2.50$\times 10^{-3}$\\
1.801 & 0.496 & 0.721 & -2.43$\times 10^{-3}$\\
1.850 & 0.536 & 0.728 & -1.46$\times 10^{-2}$\\ \hline

\textbf{1.8} & \textbf{1} & \textbf{0.72} & \textbf{0}\\
1.799 & 0.995 & 0.720 & 1.72$\times 10^{-3}$\\
1.798 & 0.988 & 0.720 & 6.72$\times 10^{-4}$\\
*1.830 & 0.990 & 0.722 & -1.74$\times 10^{-2}$\\

\end{tabular}
\end{center}
\caption{Continuation of Table~\ref{table4} ($\alpha=1.8$).}
\label{table6}
\end{footnotesize}
\end{table}


\subsection{Quartic potential\label{sec:quartic}}

For the quartic potential $V(x)=x^4/4$ ($\hat{\mathbf{V}}=k\frac{\partial^3}{\partial k^3}$) and symmetric $\alpha$-stable noises, fractional Fokker-Planck equation in the Fourier space has the form
\begin{equation}
\frac{\partial^3\hat{P}(k)}{\partial k^3}=\mathrm{sign}k|k|^{\alpha-1}\hat{P}(k).
\label{eq:quartic}
\end{equation}
The solution of Eq.~(\ref{eq:quartic}) is known for $\alpha=1$ \cite{chechkin2002,chechkin2003},
$\hat{P}_{\alpha=1}(k)=\frac{2}{\sqrt{3}}\exp\left(-\frac{|k|}{2}\right)\cos\left (\frac{\sqrt{3}|k|}{2} -\frac{\pi}{6}\right),$
and the corresponding stationary solution in the real space reads
\begin{equation}
P_{\alpha=1}(x)=\frac{1}{\pi(1-x^2+x^4)}.
\label{eq:cauchystat}
\end{equation}
The stationary solutions (\ref{eq:cauchystat}) of Eq.~(\ref{eq:quartic}) have two main properties. First of all the stationary PDFs are not of the Boltz\-mann type, as typical for the stationary states for system driven by L\'evy white stable noises with the stability index $\alpha<2$ \cite{eliazar2003}. Additionally, the stationary probability density function for a quartic Cauchy oscillator is bimodal, with extremes located at $x=\pm \nicefrac{1}{\sqrt{2}}$ \cite{chechkin2002,chechkin2003,chechkin2004}. Fig.~\ref{fig:static} presents stationary PDFs obtained by the simulation of the Langevin equation (\ref{eq:langevin}) along with theoretical lines for the Gaussian (left panel) and Cauchy (right panel) quartic oscillators. Moreover, the parabolic addition to the quartic potential $V(x)=ax^2/2+x^4/4$ $(a>0)$, can diminish or even destroy the bimodality of the stationary PDF \cite{chechkin2002,chechkin2003,chechkin2004}. Finally, for the general $\alpha$-stable driving and the quartic potential the stationary density fulfills
\begin{equation}
\frac{\partial^3\hat{P}(k)}{\partial k^3}=\mathrm{sign}k|k|^{\alpha-1}\hat{P}(k)-i\beta\tan\frac{\pi\alpha}{2}|k|^{\alpha-1}\hat{P}(k).
\label{eq:quarticf}
\end{equation}

Quartic potentials pose some additional difficulties to Monte-Carlo simulations
making it necessary to slightly modify the standard techniques of integration of
stochastic differential equations driven by $\alpha$-stable L\'evy type noise
\cite{janicki1994}. Due to heavy tails of stable distribution large random
pulses are much more likely to occur than in the Gaussian distribution leading
to very long jumps from time to time putting a particle to the position where
the force acting on it is very large. In this case approximating the
deterministic drift by $-\Delta t V'(x)$ is too inaccurate whatever small time
step is chosen. It can result in switch of the particle to the other side of the
origin in such a way that new particle's position is more distant from the
origin than the initial one leading to numerical instability and escape of the
particle to infinity \cite{dybiecphd}. Of course, this effect is weaker when
smaller integration step is used. However, taking smaller steps was proven not
to give an effective solution to the problem. Our approach to it is based on
separating noise and deterministic drift and integrating the last one
analytically, by solving the differential equation $\dot{x}(t) = - V'(x)$ to
obtain $x(t+\Delta t)$ for a given initial condition $x(t)$. Such a step
(involving the solution of an algebraic equation) is more time-consuming than
the Euler integration step and is absolutely superfluous for small and moderate
$x$. Therefore, exact integration of the deterministic part is performed only for
large $x$, $|x|>\xtr$, while the noisy part is always integrated in the standard
Euler way. In the simulations we took $\xtr=15$ as motivated by analytical
estimates and by numerical tests. For testing purposes constructed numerical
results for the Cauchy noise ($\alpha=1$) were compared with the known
analytical solution, see right panel of Fig.~\ref{fig:static}, leading to the
excellent level of agreement.

\subsection{Double well potential\label{sec:doublewell}}

\begin{figure}
\begin{center}
\includegraphics[angle=0,width=8cm]{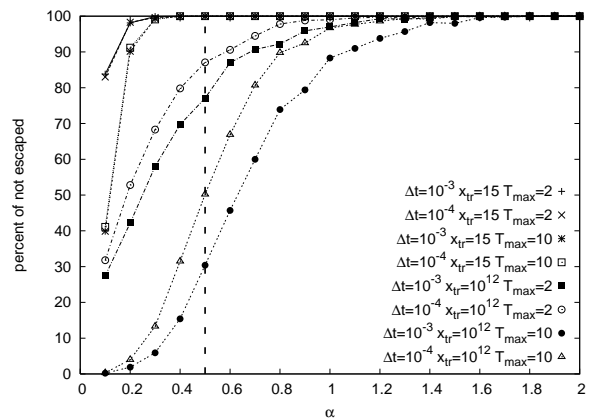}
\caption{Influence of the time step of integration and duration of simulation on portion of non escaped trajectories at time $\Tm$ for $V(x)=x^4/4-x^2/2$. $\xtr$ represents threshold value of $x$ such that for $|x|>\xtr$ deterministic part of Eq.~(\ref{eq:langevin}) is integrated analytically. The case $\xtr=10^{12}$ corresponds simply to the standard Euler scheme. For more details see the text.}
\label{fig:escape}
\end{center}
\end{figure}

\begin{figure}
\begin{center}
\includegraphics[angle=0,width=8cm]{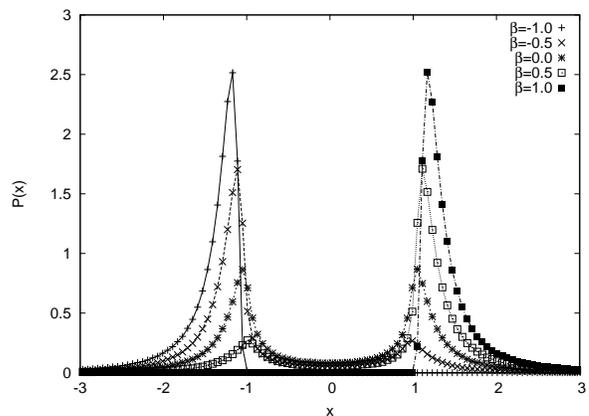}
\caption{Stationary states for the generic double well potential model subjected to the $\alpha$-stable driving with $\alpha=0.5$ and various $\beta$. Simulation parameters: $N=10^6$, $\Delta t=10^{-3}$, $\Nb=100$, $\xtr=15$ and $\Tm=10$. Stationary states for totally skewed noise, i.e., $\beta=\pm1$ are one-sided.}
\label{fig:alpha05}
\end{center}
\end{figure}

\begin{figure}
\begin{center}
\includegraphics[angle=0,width=8cm]{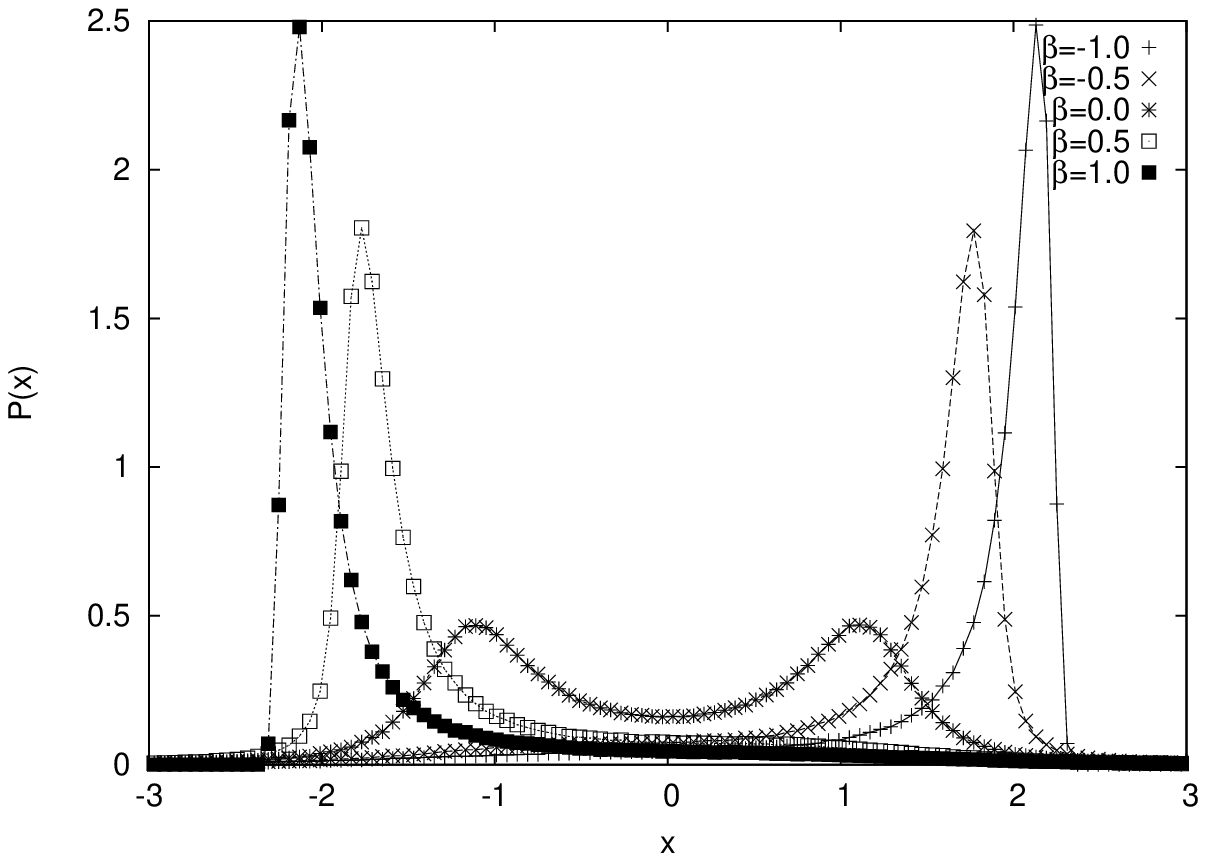}
\caption{The same as in Fig.~\ref{fig:alpha05} for $\alpha=1.1$.}
\label{fig:alpha11}
\end{center}
\end{figure}

\begin{figure}
\begin{center}
\includegraphics[angle=0,width=8cm]{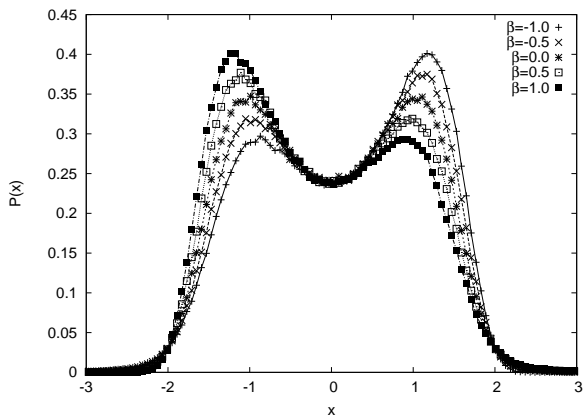}
\caption{The same as in Fig.~\ref{fig:alpha05} for $\alpha=1.8$.}
\label{fig:alpha18}
\end{center}
\end{figure}

The results for the double-well potential model, see Fig.~\ref{fig:density},  were constructed by the
numerical method described in the previous section (Section~\ref{sec:quartic}).
Furthermore, we compared the influence of decreasing time step of integration
and $\xtr$. The results of comparison of both methods of reduction of number of
numerical escapes are summarized in Fig.~\ref{fig:escape} where the influence of
the time step of integration $\Delta t$, duration of the simulation $\Tm$ on the
portion of valid (non escaped) trajectories are compared.

\begin{figure}
\begin{center}
\includegraphics[angle=0,width=8cm]{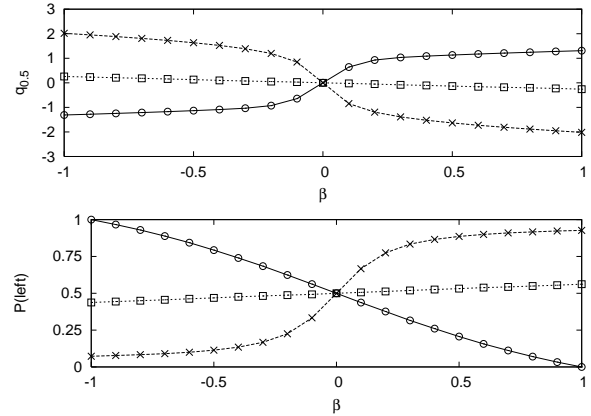}
\caption{Median (quantile $q_{0.5}$) of the stationary probability density
$P(x)$ (upper panel) and asymmetry of stationary distributions measured as the
fraction of the probability mass on the left hand side of the origin, i.e.,
$P(\mathrm{left})=\int_{-\infty}^0P(x)dx$ (lower panel). For symmetric noises
resulting stationary densities are symmetric along $x=0$ and consequently
$P(\mathrm{left})=1/2$ and $q_{0.5}=0$. Different symbols correspond to
different values of the stability index $\alpha$: `$\circ$' $\alpha=0.5$,
`$\times$' $\alpha=1.1$ and `$\square$' $\alpha=1.8$.}
\label{fig:splitting}
\end{center}
\end{figure}

Stationary solutions shown in Figs.~\ref{fig:alpha05}--\ref{fig:alpha18} are
obtained for the generic double well potential model, i.e., $V(x)=x^4/4-x^2/2$.
In the simulation the whole allowed range of $\alpha$ and $\beta$ was examined,
in figures only a limited choice of representative values of noise parameters is
presented, namely the same as ones used in Tabs.~\ref{table4}--\ref{table6}. For
$\alpha<1$ with $|\beta|=1$ stable distributions are one-sided, therefore,
stationary solutions for $\alpha=0.5$ with $\beta=\pm1$ are different from zero
only on the one side of the origin. For $|\beta|<1$ stable distributions takes
all real values as manifested by nonzero probability for all $x$, see
Fig.~\ref{fig:alpha05}. Furthermore, the symmetry of noise and the potential is
reflected in the symmetry of stationary densities, i.e., solutions for $-\beta$
can be constructed by the reflection of the solutions for $\beta$, see
Figs.~\ref{fig:alpha05}--\ref{fig:alpha18}. For $\beta=0$ with any $\alpha$
stationary densities are bimodal and symmetric along $x=0$. For $\alpha>1$, the
support of stable densities as well as the stationary distributions is whole
real line. Consequently, even extreme values of the asymmetry parameter
$\beta=\pm1$ are not sufficient to switch the probability mass to the one side
of the origin, see Figs.~\ref{fig:alpha11}--\ref{fig:alpha18}. Furthermore, it
is well documented in the lower panel of Fig.~\ref{fig:splitting} where
$P(\mathrm{left})=\int_{-\infty}^0P(x)dx$ is presented. Theoretical
considerations as well as the probability of being in the left state, $P(\mathrm{left})$, indicate that
stationary densities can be one-sided only for some totally skewed
$\alpha$-stable noises with small $\alpha$, i.e., $\beta=\pm1$ with $\alpha \ll 1$. For
$\alpha>1$ with $\beta=\pm1$ two maxima of stationary PDFs are visible. In upper
panel of Fig.~\ref{fig:splitting} locations of the median value, which may be
considered as the next measure of asymmetry of stationary probability densities,
are presented.

Very small values of $\alpha$ ($\alpha<0.5$) pose special difficulties for simulations.
Our simulation of Eq.~(\ref{eq:langevin}) starts with the initial condition $x(0)=0$.
The initial transient peak of the probability density at this value is rather persistent
for small $\alpha$, so that the simulation time has to be long.
On the other hand in this case simulations are prone to escape of trajectories to ``infinity''
due to too strong noise pulses and require very small $\Delta t$, so that the overall quality of
such results is not very good. Therefore, the results for $\alpha < 0.5$ are not presented here.

\begin{figure} \begin{center}
\includegraphics[angle=0,width=8cm]{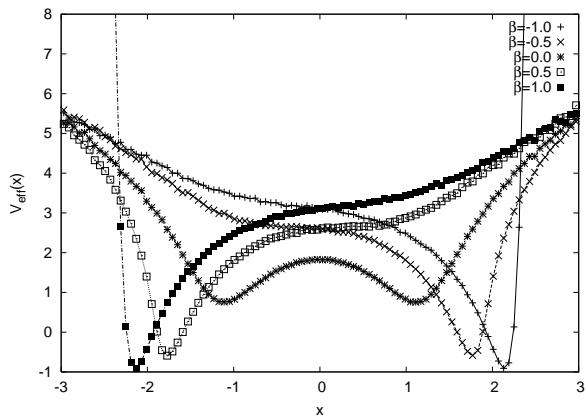} \caption{Effective
potentials $\Veff=-\ln P(x)$ corresponding to stationary densities from
Fig.~\ref{fig:alpha11}. The effective potential $\Veff$ together with the white
Gaussian noise results in stationary densities of the model (\ref{eq:langevin})
which are presented in Fig.~\ref{fig:alpha11}.}
\label{fig:alpha11_eff}
\end{center}
\end{figure}

Another method to present the results for stationary distributions is the use of
the effective potential. In general, the same stationary probability densities
that are recorded for the double well system driven by $\alpha$-stable noise,
see Eq.~(\ref{eq:langevin}), can be observed in the Gaussian regime in the
effective potential $\Veff=-\ln P(x)$. Sample effective potentials corresponding
to stationary states for $\alpha=1.1$ from Fig.~\ref{fig:alpha11} are depicted
in Fig.~\ref{fig:alpha11_eff}. The same stationary solutions can be observed for
motion in a simple potential, like double well potentials, and $\alpha$-stable
stochastic driving or for potentials of the complicated form, see
Fig.~\ref{fig:alpha11_eff}, and standard white Gaussian driving. Despite the
fact that stationary states are the same, other characteristics of these two
processes are different.

\section{Summary and conclusions\label{sec:summary}}

In the present work we investigated the form of stationary probability densities of
a position of a particle subject to a deterministic potential force and to
a L\'evy noise, paying special attention to the case of asymmetric stable noises.
Stationary density functions for system driven by $\alpha$-stable
L\'evy noises ($\alpha \neq 2$), if they exist, are not of the Boltzmann-Gibbs form.

For parabolic potential the stationary density functions are
$\alpha$-stable laws with the same stability index $\alpha$ and
asymmetry parameter $\beta$ as ones of the noise. The only
difference is the scale parameter of the resulting distribution.
Therefore, this case can serve as a benchmark for our simulation
algorithms. By use of the special software \cite{stable} for estimation of
stable law parameters, the parameters of stationary densities have been
evaluated leading to a very good level of agreement between
theoretical and estimated values of parameters. Natural
consequence of the L\'evy type of the stationary densities for
parabolic potential is divergence of the variance of the particle
position. To produce bounded states, i.e., states with finite
variance of a position, potentials steeper than quadratic are necessary.
Thus, for the quartic potential variance of stationary
densities is finite. Furthermore, for white Cauchy noise
analytical solutions to the stationary problem is known. Here
again, numerical results fully agree with earlier
\cite{chechkin2002,chechkin2003} theoretical findings.

In the case of Gaussian noise ($\alpha=2$) the symmetry of stationary density
(being the Boltzmann-Gibbs equilibrium distribution) reflects
symmetries of the underlying potential: the asymmetric densities correspond
to asymmetric potentials, i.e., to deterministic forces which break the symmetry of the initial
problem. The Gaussian noise itself is always symmetric.
For systems driven by L\'evy noises ($\alpha \neq 2$), an asymmetric stable noise together
with symmetric static potential is sufficient to produce asymmetric
stationary densities. In this situation the asymmetry of stationary states originates
from the asymmetry of the stochastic driving and can be controlled by changing the parameters
of the noise.

Our main studies have been performed for the generic symmetric double well
potential model. Here the asymmetry of the stationary distribution (as measured
by the probability of being in the left/right state or location of the median)
was investigated as a function of the parameters of the noise. The asymmetry of
stationary state decreases with increasing $\alpha$, see
Fig.~\ref{fig:splitting}. Finally for $\alpha=2$ with any value of the asymmetry
parameter $\beta$ Gaussian scenario is recovered and stationary density is fully
symmetric. We also checked whether stable asymmetric noise can produce unimodal
stationary PDFs in the double well potential. Such situation can indeed be
observed for fully asymmetric noises ($|\beta|=1$) with $\alpha \ll 1$, e.g. for
the L\'evy-Smirnoff noise ($\alpha=0.5$, $\beta=1$).

\appendix
\section{Dimensionality of the Langevin equation\label{sec:appendix}}
We have investigated the dynamic stochastic process modeled by the (overdamped) Langevin equation of the form
\begin{equation}
\dot{x}(t)=\frac{f(x)}{\gamma m} + D^{1/\alpha}\zeta(t)=\frac{f(x)}{\gamma m} + \sigma\zeta(t),
\label{eq:langlwn}
\end{equation}
where: $x$ -- is a position of the particle, $\gamma$ -- stands for a friction coefficient, $m$ -- is particle's mass, $D$ ($\sigma$) -- represents strength of the noise and $\zeta(t)$ -- is L\'evy stable white noise characterized by the stability index $\alpha$ ($\alpha\in(0,2]$) and the asymmetry parameter $\beta$ ($\beta\in[-1,1]$). The force acting on a particle is determined by the external potential, $f(x)=-dV(x)/dx$.

Corresponding units in Eq.~(\ref{eq:langlwn}) are: $[x]=[\mbox{length}]$, $[\gamma]=1/[t]$, $[f(x)]=[V'(x)]=[m]\times[\mbox{length}]/[t]^2=[\mbox{force}]$, $[V(x)]=[\mbox{force}]\times[\mbox{length}]=[\mbox{energy}]$, $[D]=[\mbox{length}]^\alpha/[t]$ ($[\sigma]=[\mbox{length}]/[t]^{1/\alpha}$) and $[\zeta(t)]=1/[t]^{1-1/\alpha}$. Stability index $\alpha$ and asymmetry parameter $\beta$ are dimensionless. In the asymptotic limit of $\alpha=2$ the L\'evy white noise is equivalent to the Gaussian white noise and it has standards units, i.e. $[\zeta_{\alpha=2}(t)]=[\xi(t)]=1/\sqrt{[t]}$.

By the set of transformation:
$t \to t/t_0$ and
$x\to x/x_0$.
Eq.~(\ref{eq:langlwn}) can be transformed to the dimensionless form
\begin{equation}
\dot{x}(t)=f(x) + D^{1/\alpha}\zeta(t)=f(x) + \sigma\zeta(t),
\label{eq:langlwndl}
\end{equation}
which is of the same type like Eq.~(\ref{eq:langevin0}) because the (rescaled) noise intensity can be incorporated to the distribution of the particle's position increments.
An alternative way of getting dimensionless form of Eq.~(\ref{eq:langlwn}) can be found in a recent work \cite{chechkin2007}.

For the single-well potential $V(x)=ax^n/n$
\begin{equation}
t_0=\frac{x_0^\alpha}{D},
\;\;\;\;\;
x_0=\left[ \frac{D \gamma m}{a}\right] ^{\frac{1}{n-2+\alpha}}.
\end{equation}
Thus
$V(x) \to x^n/n$ and
$\sigma \to 1$.
Therefore, for the single minima potential, the only relevant parameters are $\alpha$ (stability index) and $\beta$ (asymmetry parameter).

For the generic double-well potential $V(x)=-ax^2/2+bx^4/4$
\begin{equation}
t_0=\frac{\gamma m}{a},
\;\;\;\;\;
x_0^2=\frac{\gamma m}{bt_0}=\frac{a}{b}.
\end{equation}
Thus
$V(x)\to-x^2/2+x^4/4$ and
$\sigma \to  \sigma t_0^{1/\alpha}/x_0$.
In consequence, for the double-well potential, the only relevant parameters in a dimensionless form of Eq.~(\ref{eq:langlwn}) are the L\'evy noise parameters $\alpha$ and $\beta$,  and  a  (rescaled) noise strength. From the above analysis, it is clear that the choice of $x_0$, $t_0$ introduce
scales to the system which are directly related to its dynamical parameters.

\begin{acknowledgments}
The research has been supported by the Marie Curie TOK COCOS grant (6th EU Framework Program under
Contract No. MTKD-CT-2004-517186). Computer simulations have been performed at
the Academic Computer Center CYFRONET AGH, Krak\'ow. \ Additionally, BD
acknowledges the support from the Foundation for Polish Science
 and the hospitality of the Humboldt University of Berlin and the Niels Bohr Institute (Copenhagen).
 The support by DFG within SFB 555 is also acknowledged.
\end{acknowledgments}


\end{document}